\documentstyle[12pt,epsf,epsfig]{article}  
\def\li2{{\rm Li}_2}

\def\roughly#1{\,\,\raise.3ex\hbox{$#1$\kern-.75em\lower1ex\hbox{$\sim$}}\,\,}

\def \gsim{\mathrel{\vcenter
     {\hbox{$>$}\nointerlineskip\hbox{$\sim$}}}}

\def\fo{\hbox{{1}\kern-.25em\hbox{l}}}

\def\bea{\begin{eqnarray}}
\def\eea{\end{eqnarray}}
\def\beq{\begin{equation}}
\def\eeq{\end{equation}}
\def\eq{\end{equation}}
\def\to{\rightarrow}

\def\bsg{\ifmmode B\to X_s\gamma\else $B\to X_s\gamma$\fi}
\def\bsll{\ifmmode B\to X_s\ell^+\ell^-\else $B\to X_s\ell^+\ell^-$\fi}
\def\bstt{\ifmmode B\to X_s\tau^+\tau^-\else $B\to X_s\tau^+\tau^-$\fi}
\def\shat{\ifmmode \hat{s}\else $\hat{s}$\fi}

\newcommand{\newc}{\newcommand}

\newc{\lcal}{\int {\cal L}dt}
 
\newc{\LSP}{{\chi^0_1}}
\newc{\stauR}{{\tilde \tau_R}}
\newc{\stau}{{\tilde \tau_1}}
\newc{\mstop}{m_{\tilde{t}}}
\newc{\mHpm}{m_{H^\pm}}
\newc{\ie}{{\it i.e.}}          
\newc{\etal}{{\it et al.}}
\newc{\eg}{{\it e.g.}}          
\newc{\kev}{\hbox{\rm\,keV}}            
\newc{\mev}{\hbox{\rm\,MeV}}            
\newc{\gev}{\hbox{\rm\,GeV}}            
\newc{\tev}{\hbox{\rm\,TeV}}
\newc{\xpb}{\hbox{\rm\, pb}}
\newc{\xfb}{\hbox{\rm\, fb}}

\newc{\mtop}{m_t}
\newc{\mbot}{m_b}
\newc{\mz}{m_Z}
\newc{\mw}{M_W}
\newc{\alphasmz}{\alpha_s(m_Z^2)}
\newc{\swsq}{\sin^2\theta_W}
\newc{\tw}{\tan\theta_W}
\newc{\cw}{\cos\theta_W}
\newc{\sw}{\sin\theta_W}
\newc{\BR}{\hbox{\rm BR}}
\newc{\zbb}{Z\to b\bar}
\newc{\Gb}{\Gamma (Z\to b\bar b)}
\newc{\Gh}{\Gamma (Z\to \hbox{\rm hadrons})}
\newc{\rbsm}{R_b^\hbox{\rm sm}}
\newc{\rbsusy}{R_b^\hbox{\rm susy}}
\newc{\drb}{\delta R_b}

\newc{\sgn}{\mbox{sgn}}
\newc{\tbeta}{\tan\beta}
\newc{\uL}{{\tilde u_L}}
\newc{\uR}{{\tilde u_R}}
\newc{\cL}{{\tilde c_L}}
\newc{\cR}{{\tilde c_R}}
\newc{\tL}{{\tilde t_L}}
\newc{\tR}{{\tilde t_R}}
\newc{\dL}{{\tilde d_L}}
\newc{\dR}{{\tilde d_R}}
\newc{\sL}{{\tilde s_L}}
\newc{\sR}{{\tilde s_R}}
\newc{\bL}{{\tilde b_L}}
\newc{\bR}{{\tilde b_R}}
\newc{\eL}{{\tilde e_L}}
\newc{\eR}{{\tilde e_R}}
\newc{\mhp}{m_{H^\pm}}
\newc{\mhalf}{m_{1/2}}
\newc{\emt}{{e/\mu /\tau}}

\newc{\bW}{{\bar W}}
\newc{\bB}{{\bar B}}
\newc{\eps}{{\epsilon}}

\def\lappeq{\mathrel{\rlap{\raise.5ex\hbox{$<$}}
{\lower.5ex\hbox{$\sim$}}}}
\def\gappeq{\mathrel{\rlap {\raise.5ex\hbox{$>$}}
{\lower.5ex\hbox{$\sim$}}}}

\begin{document}

 


\vspace{20pt}
\font\cmss=cmss10 \font\cmsss=cmss10 at 7pt 



\hfill

\vspace{20pt}

\begin{center}
{\Large \textbf
{ElectroWeak Symmetry Breaking as of 2003, on the way to the Large Hadron Collider} \footnote{Lectures given at the Cargese School of Physics and Cosmology - August 2003 - Cargese - France} }
\end{center}

\vspace{6pt}

\begin{center}
\textsl{$R. Barbieri$}\vspace{20pt}

\textit{Scuola Normale Superiore and INFN, Pisa, Italy}

\end{center}

\vspace{12pt}

\begin{center}
\textbf{Abstract }
\end{center}

\vspace{4pt} {\small \noindent 
I review the status of the ElectroWeak Symmetry Breaking problem. The lectures are naturally divided into two parts. The first is mostly devoted to overview  the impact of current data on the issue of EWSB. The tools are known, the latest data are included. Always in the first part, I say why I care about the "little hierarchy" problem and I summarize how some proposals for EWSB, recent and less recent, are confronted with this problem. Motivated by these considerations, in the second part I describe the essential features of a proposal for breaking supersymmetry, and consequently the electroweak symmetry, by boundary conditions on an extra dimension.}

\vfill\eject 
\noindent


\section{The data (their interpretation) summarized}
\label{sec:data}

There are several good reasons for being interested in the problem of how the electroweak symmetry gets broken. Above all, the physical origin of the Fermi scale has not been identified, yet. Consequently, and not less importantly, this ignorance acts as a cloud on every attempt to design a theory of the fundamental interactions beyond the Standard Model (SM). Last, but not least, the exploration of the TeV scale of energy expected at the Large Hadron Collider (LHC) should finally allow a direct comparison with experiment of every theoretical idea on this matter.

These are not the first lectures on the subject of ElectroWeak Symmetry Breaking (EWSB). Nevertheless, I find that it may be useful to overview the present status of the subject when we still have a few years before the start of the LHC and when the program of the ElectroWeak Precision Tests (EWPT), in particular with the completion of most of the data analyses by the LEP experiments, is in a mature stage.

The EWPT are still the most important source of experimental information, although indirect, on EWSB. It so happens that I already lectured in Cargese on the EWPT in 1992, when the accumulation of significant experimental results on the EWPT was about to start and the top quark had not yet been discovered. At that time the bulk of the radiative effects seen in the data was still of electromagnetic origin. Now we know that several per-mil effects of pure electroweak nature are crucial in allowing an effective description of the data and that these effects are contained in the SM. Things might have gone differently.

What is it then that we learn on the EWSB problem? Among the conclusions of my 92 lectures, I argued that the program of the EWPT should have made possible to discriminate between a perturbative and a strongly interacting picture of EWSB, the prototype examples for the two cases being respectively supersymmetry and technicolor. It is now pretty clear that the data support a perturbative more than a non perturbative description of EWSB, as illustrated more precisely later on. Inside this framework, a relevant piece of information, also coming from the EWPT is the indication for a light Higgs, most likely lighter than about 200 GeV. All this seems in fact to make a rather coherent picture of EWSB, and maybe it does. I will argue, however, that the direct lower limit on the Higgs mass, $m_H > 115 \, GeV$ \cite{Higgssearch}, and the absence so far on any deviation from the expectations of the SM may also require some interpretation with a possible impact on the picture of EWSB. 

The lectures are naturally divided into two parts, to be found in Sections 2 and 3 respectively. The first is mostly devoted to overview the impact of current data on the issue of EWSB. The tools are known, the latest data are included. Always in the first part, I will illustrate why I care about the "little hierarchy" problem and I will summarize how some proposals for EWSB, recent and less recent, are confronted with this problem. Motivated by these considerations, in the second part I will describe the essential features of a proposal for breaking supersymmetry, and consequently the electroweak symmetry, by boundary conditions on an extra dimension. In Section 4 I summarize the line of reasoning that motivates mostly these lectures and I conclude.

\section{Un updated overview}
\label{overview}

\subsection{The data (their interpretation) summarized}

\subsubsection{Experiment versus theory with generic "oblique" corrections}
\label{Exp-ver-theo}

I begin by referring to the data on the EWPT from the LEP, TEVATRON and SLC experiments, as summarized  by the LEP ElectroWeak Working Group in the summer of 2003 \cite{lep}. These data allow a stringent test of the SM, sensitive to the radiative corrections of electroweak nature. The test is successful, with no serious reason of concern, in my view, for those measurements that appear in some tension with the SM prediction. 

\begin{figure}
  \centering \includegraphics[width=10cm]{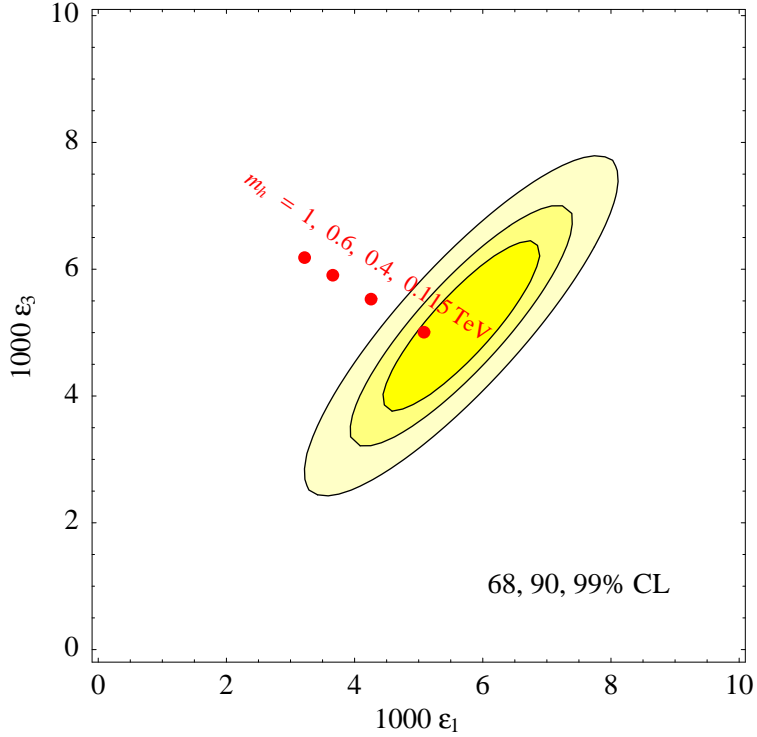}
   \caption{$\epsilon_1$ and $\epsilon_3$, with their correlation, as determined from the EWPT. The SM prediction is indicated for different values of the Higgs mass.}
   \label{fig:eps1eps3}
\end{figure}

The EWSB sector of the SM has some impact on this test. For this very reason, taking into account that this is the physics  whose nature we are wondering about, one has advocated since the beginning of the experimental program an analysis valid in a broader class of theories. Such are those theories that differ from the SM only in the so called "oblique" corrections  \cite{pt2}, i.e. those corrections that come from vacuum polarization amplitudes of the vector bosons. To this purpose one has defined three dimensionless experimental quantities, $\eps_i, i=1,2,3$, \cite{alba} with the property that they encapsulate, among other effects, the "oblique"  corrections. Furthermore, since one of them, $\eps_2$, is unlikely to contain new physics, one often freezes it to its SM value \footnote{The slight dependence of  $\eps_2$ on $m_H$ is practically irrelevant. Here $\eps_2$ is taken at $m_H = 115 \, GeV$.}. Fig. \ref{fig:eps1eps3} shows the determination of the two remaining parameters, $\eps_1$ and $\eps_3$, with their correlation, as obtained from the set of data mentioned above. In the same Figure, the SM prediction is indicated for different values of the Higgs mass.

The "oblique" contributions to $\eps_1$ and $\eps_3$ can be expressed in terms of the transverse components of the usual vacuum polarization amplitudes $\Pi_{a, b}(q^2)$, with $a, b = 1, 2, 3$ for the SU(2) or $a, b = B$ for the U(1) gauge bosons, as
\beq
\Delta \epsilon_1 = \frac{ \Pi_{33}(0) - \Pi_{11}(0)} {M_W^2},
\eeq
\beq
\Delta \epsilon_3 = \frac{ g} {g'} \Pi_{3B}'(0),
\eeq
where $g$ and $g'$ are the SU(2) and U(1) gauge couplings.

The data speak by themselves. The agreement with the SM  is remarkable and constitutes indirect evidence for the existence of the Higgs. It has even become difficult, if not impossible at all,  to try to reconcile with the data a heavy Higgs, in the TeV range of mass where the very concept of the Higgs becomes elusive, by adding an extra negative contribution to $\eps_3$.  As commented below, a positive extra effect on $\eps_1$, or a combination of two effects, would appear more useful in this respect. On the other hand, and always with a heavy Higgs, a positive extra contribution to $\eps_3$ above  2 units in $10^{-3}$ is excluded, no matter what $\eps_1$ does. If there is a Higgs, which seems more likely, Fig. \ref{fig:eps1eps3} makes also clear the preference of the EWPT for  a low mass value, close to the direct lower bound of 115 GeV.

\subsubsection{Constraining the cut-off of the SM}
\label{constr-cutoff}

An alternative way to try to appreciate the impact of the EWPT is to view the Lagrangian of the SM as an effective low energy theory with possible modifications introduced by operators ${\cal O}_i^{(4+p)}$ of dimension $(4+p)$,
\beq
{\cal L}_{eff}(E < \Lambda)= {\cal L}_{SM} + \sum_{i,p} \frac{c_i}{\Lambda^p}{\cal O}_i^{(4+p)}.
\label{gen SM}
\eeq
How does this modified Lagrangian compare with the most recent EWPT \cite{bar-str}? Table \ref{limits} gives a list of some  gauge invariant operators, together with the lower limit that the same EWPT set on the scale $\Lambda$ that multiplies each of them. One operator at a time is taken, with the dimensionless coefficients $c_i=+1$ or $c_i=-1$ (the interference with the SM amplitude matters) and the Higgs mass (which affects the SM amplitude) at 115 GeV. 

   \begin{table}
  \centering
    \begin{tabular}{|c|c|c|}
\hline
\hline
     Dimension six operator &  $c_i = - 1$ &  $c_i = + 1$  \\
\hline
    $ {\cal O}_{W B} = (H^+ \sigma^{a} H) W_{\mu \nu}^a B_{\mu \nu}$ & 9.0 &  13 \\
\hline
    $ {\cal O}_{H} = \vert H^+ D_{\mu} H) \vert ^2 $ & 4.2 &  7.0 \\
\hline
    $ {\cal O}_{LL} = \frac{1}{2} (\bar{L} \gamma_{\mu} \sigma^a L)^2 $ & 8.2 &  8.8 \\
  \hline
    $ {\cal O}_{HL} = i (H^+ D_{\mu} H) (\bar{L} \gamma_{\mu} L) $ & 14 &  8.0 \\
     \hline
    \end{tabular}

  \caption{$95\%$ lower bounds on $\Lambda/TeV$ for the individual operators (and the coefficients $c_i$ as indicated) with $m_H = 115 \, GeV$. $\sigma^a$ are the Pauli matrices acting on the $SU(2)_L$ doublets.}
  \label{limits}
\end{table}

Table \ref{limits} gives a feeling on the possible lowest energy at which a change of regime can intervene relative to the SM physics, called   the "cut-off " of the SM. To be able to make a more precise statement, the coefficients $c_i$ would have to be known. On general grounds, the most important conclusion from Table 1 is that, if a new strong interaction intervenes at $\Lambda$, it is unlikely that this scale be lower than about 10 TeV. A different statement might be defendable, but only on a case by case basis. 

In the case of the operators ${\cal O}_{WB}$ and ${\cal O}_H$, but only in these cases, there is a direct connection of Table 1 with Fig. \ref{fig:eps1eps3}, since these operators affect the EWPT only by contributing to the $\epsilon$-coefficients, via

\beq
\delta \epsilon_3 ({\cal O}_{WB})= \frac{2 g}{g'} c_{WB} \frac{v^2}{\Lambda^2},
\eeq
\beq
\delta \epsilon_1 ({\cal O}_H)= -  c_H \frac{v^2}{\Lambda^2}.
\label{epsilon_1}
\eeq
It is also clear therefore that a theory capable of modifying the SM by the  inclusion of a single dominant operator ${\cal O}_H$ would allow a fit to the EWPT with a heavier Higgs (with $c_H=-1$ and a corresponding scale lower than in Table \ref{limits}) \cite{Hall:1999fe, bar-str}.

\subsection{The "little hierarchy" problem }
\label{hierarchy}

It is generally believed by many, including myself, that the discovery of the Higgs would not identify, \emph {per se}, the physical origin of the 
Fermi scale. It is also a widespread opinion that the LHC experiments will reveal new phenomena related to EWSB and not included in the SM. Let us review the argument that supports this view. The general argument is old. Its numerical terms are new inasmuch as one uses in an essential way the precise knowledge of the top mass and the indirect information on the Higgs mass, as just described.

For a fixed value of the Higgs vacuum expectation value (vev) or of the Fermi scale, the physical Higgs mass can either be computed from the quartic Higgs coupling or from the curvature of the Higgs potential at zero field. Using this last quantity, the one loop radiative correction to the squared Higgs mass in the SM is

\beq
\delta m_H^2 =   \frac{6 G_F}{\sqrt{2} \pi^2} (m_t^2 - 1/2 m_W^2 - 1/4 m_Z^2 - 1/4 m_H^2)
 \int^{\Lambda}E dE,
 \label{delta m_H^2}
\eeq
where $E=\sqrt{k^2}$, $k$ is the momentum running in the loop, and terms of relative order $(m/\Lambda)^2$ for any particle mass are neglected. For the cut-off of the integral, taken equal for all the individual contributions, the use of the same symbol as the one employed for the cut-off of the SM is not  without reason.
As apparent from eq.~(\ref{delta m_H^2}),
if one accepts that the Higgs mass is below 200 GeV, as I do in the following unless otherwise stated, the top loop contribution, proportional to the top mass squared, is clearly dominant. Furthermore it becomes numerically significant for a pretty low value of $\Lambda$, since, using $m_t = 174 GeV$,
\beq
\delta m_{H,top}^2(SM) \sim  ( 115 GeV)^2  (\frac{\Lambda}{400 GeV} )^2 .
 \label{delta m_H,top}
\eeq
It will be useful for later use to define a kind of "Higgs mass spectral function" $\Delta m_H^2(E)$ which enters the expression of the radiative Higgs squared mass as
\beq
\delta m_{H}^2 \equiv   \int^{\Lambda} \frac{dE}{E} \Delta m_H^2(E).
 \label{Delta m_H}
\eeq
In the SM, $\Delta m_H^2(E)$  grows quadratically with $E$, it becomes rapidly large and it gives  the main contribution to the Higgs mass for $E$ close to the cut-off where it cannot be trusted.

Eq.~(\ref{delta m_H,top}) is the main argument for expecting new physics to show up at the LHC. How can it possibly be that $\Delta m_H^2(E)$ keeps growing unmodified up to energies unaccessible to the LHC, when already at 400 GeV the radiative correction to the Higgs mass is around the best value implied by the EWPT as discussed above? Note that this is not an inconsistency of the SM. Since the curvature of the Higgs potential is a free parameter of the SM Lagrangian, it only takes a little counterterm to subtract away the loop effect that we are discussing.  I prefer to think - or to assume, or to hope, as one prefers - that the Higgs mass is what it is not as a  result of a \emph {fortuitous} cancellation between $\delta m_{H,top}^2 $ and the counterterm or any other effect. For this to be the case, a  physical mechanism must prevent $\Delta m_H^2(E)$ from growing. In turn, this physical mechanism, not included in the SM, should give effects observable at the LHC.

Ideally, one would like to be quantitative here and be able to estimate at which energy these new phenomena should show up, since this is likely to  be crucial for a test at the LHC. This is possible in specific models and, if it is the case, for a given amount of fine tuning that one is willing to tolerate. On general grounds, however, it is possible to say that the physical mechanism that dumps $\Delta m_H^2(E)$ has to pass a nontrivial test: it must not  disturb the agreement of the SM with the EWPT, as summarized in the previous Section. To find such a mechanism, involving a minimum amount of tuning among otherwise uncorrelated parameters, is to solve the "little hierarchy" problem \cite{rb}. The hierarchy is between the Higgs mass, supposedly close to $100 \, GeV$, and the lower limit on the cut-off of the SM, as elaborated upon always in the previous Section.

\subsection{Some ideas on EWSB with a light Higgs }
\label{ideas-on-EWSB}

In this overview, I find it useful to comment on two proposal for EWSB, since they may accommodate or even imply a light Higgs, while addressing at the same the "little hierarchy" problem. One is EWSB triggered by supersymmetry breaking, as in the Minimal Supersymmetric Standard Model (MSSM).  The other assumes some new strong force which gives rise to a pseudo-Goldstone boson with the quantum numbers of the standard Higgs, supplemented with the mechanism that goes under the name of "little Higgs".  A general overview of these two schemes goes beyond the scope of these lectures. Rather I concentrate my attention on the way they tackle the "little hierarchy" problem.

\subsubsection {The Minimal SuperSymmetric Standard Model}
\label{MSSM}

The MSSM neatly implies a light Higgs and a perturbative dynamics of EWSB.
 The Higgs is light because the quartic Higgs coupling is a gauge coupling.
 The perturbative dynamics gives, in principle, the easiest way to solve the little
 hierarchy problem. The MSSM 
 has indeed no special problem in preserving 
 the success of the SM in describing the EWPT. The effort that has gone
 and will possibly go in MSSM studies appears therefore well justified. The
 most important question nevertheless remain.
 Where are the supersymmetric particles? Even more concretely, 
 which are the chances that they will be seen at the LHC? The possible answer, 
 at the present state of knowledge, goes inevitably back to the fine tuning problem 
 in the curvature of the Higgs potential \cite{Barbieri:1987fn}.
 
 As well known, the MSSM involves two Higgs doublets. For the present purposes, 
 however, there is no essential loss in 
 assuming that the vev taken by one of them,
 the one that couples to the up quarks, is far bigger than the vev of the other, i.e.,
 in the familiar notation, $tan \beta \gg 1$. In this case the previous discussion 
 applies unaltered and we can focus, as before, on $\Delta m_{H, top}^2(E)$. 
 
 The way taken by the MSSM in dumping $\Delta m_H^2(E)$ is almost too well known
 to be recalled here. In the case of $\Delta m_{H, top}^2(E)$, the exchange of the stops freeses 
 it when E crosses
 their masses and keeps it constant up to the scale -  call it $\Lambda_{SB}$ -
 where these masses are generated
 by the specific mechanism of supersymmetry breaking and supersymmetry breaking
 transmission.
 In this way, in logarithmic approximation, $\delta m_{H,top}^2 $ is modified into \footnote {This, as the one that follows, 
 is a symplified discussion. To the least, when the energy range of
 interest is large compared to the Fermi scale, the influence on the stop masses
 of the gluino exchange is significant and has to be taken into account.}
  \beq
\delta m_{H,top}^2(MSSM) \sim  \frac{6 G_F m_t^2}{\sqrt{2} \pi^2} m_{ST}^2 log \frac{\Lambda_{SB}}{m_{ST}} \sim 0.15 \, m_{ST}^2 log \frac{\Lambda_{SB}}{m_{ST}} ,
 \label{delta m_H,SS}
\eeq
where $m_{ST}$ is a suitable average  of the stop masses.
 
The generalization of this argument calls for the entire spectrum of superpartners
in the generic region of the Fermi scale. Can one be more precise on this? The main limitation
comes from the experimental limit on the Higgs mass, $m_H > 115$ GeV, well above the Z mass,
 which is the theoretical upper bound on $m_H$ at tree level in the MSSM, 
 as determined from the quartic coupling. Fortunately this bound is invalid after radiative
 corrections, which are logarithmically sensitive to the stop masses
 and can be sizeable if $m_{ST}$ is big enough. But a big $m_{ST}$ brings back the
 problem in eq.~(\ref{delta m_H,SS}) , since $\delta m_{H,top}^2$ grows quadratically with it.
 Recent studies \cite{fine-tun} in the complex space of the MSSM parameters
 in its supergravity version \cite{MSSM} (but the situation in gauge-mediated models \cite{GGM} is not better) confirm this simple argument and set in at least a factor of 20
 the cancellation required against the term in eq.~(\ref{delta m_H,SS}) - or its counterpart -
 with a favorable choice of various parameters ( $A_t$, $tan \beta$, etc.). Ref. \cite{Giusti:1998gz}, as updated in \cite{Giudice:2003pj}, has a rather striking pictorial representation of this problem.
 
 The current limits on generic sparticle masses from direct searches or from
 the EWPT, as mentioned, are somewhat less constraining. The portion of parameter
 space of the MSSM explorable at the LHC by direct sparticle production,
 a primary goal of the machine,
 is still sizeable.  
 Nevertheless, in the spirit of the
 argument discussed above, the need of a significant cancellation in the curvature
 of the Higgs potential due to the present limit on the Higgs 
 mass  weakens any statement about where supersymmetry
 should be found, which is unfortunate. This may be irrelevant, with the MSSM around the corner of the parameter space explored so far. Alternatively it may  indicate the necessity to go beyond the MSSM \footnote{To correct the Higgs mass prediction at tree level by suitable F-term \cite{Haber:1986gz} or D-term \cite{Batra:2003nj} effects or by higher dimensional operators \cite{CEH} are examples in this direction.} or, more drastically, the need of some variation
 on the current picture of supersymmetry breaking.

\subsubsection{The little Higgs }
\label{littHiggs}

A neat way to have a light scalar, i.e. a candidate Higgs, from a strong dynamics is to arrange its symmetries and the pattern of symmetry breaking in such a way that this scalar comes out to be a pseudo-Goldstone boson, like a $\pi$ in QCD \cite{pseudoG}. The prefix \emph{pseudo} is mandatory since a genuine Goldstone could neither get a mass nor it could have a Yukawa coupling to the fermions. Both could arise, however, from a "weak" explicit breaking of the original global symmetry.

Let us concentrate on the top Yukawa coupling which is the major source of the growing $\Delta m_H^2(E)$, as defined in Sect.~\ref{hierarchy}. If this coupling between the top and the Goldstone boson were directly introduced in the Lagrangian, no change would have occurred in 
$\Delta m_{H, top}^2(E)$ relative to the SM up to the scale $\Lambda$ of the strong dynamics.  Based on Sect.~\ref{Exp-ver-theo}, this could hardly be considered, however, as a satisfactory solution of the "little hierarchy" problem: $\Delta m_{H, top}^2(E)$ needs to be cut-off well before getting to such a scale. In "little Higgs" models \cite{littleH} this is achieved by adding another  top-like quark which also couples to the Goldstone-Higgs in a globally symmetric way but has a non-symmetric, although gauge invariant, mass term. In this way, and without any adjustment of uncorrelated parameters, when $E$ crosses the mass of the heavy top, $m_T$, $\Delta m_{H, top}^2(E)$ gets frozen  and stays constant up to  $\Lambda$, where it eventually dies out. Consequently, up to non-logarithmic terms, eq.~(\ref{delta m_H,top}) is replaced by 
 \beq
\delta m_{H,top}^2(LH) \sim  \frac{6 G_F m_t^2}{\sqrt{2} \pi^2} m_T^2 log \frac{\Lambda}{m_T}.
 \label{delta m_H,little}
\eeq

This is a significant step forward in the direction of solving the "little hierarchy" problem. With some caveats, though. Le me denote by $f$ the "pion decay constant" of the strong dynamics at  $\Lambda$, so that  $\Lambda \sim 4 \pi f$. It is $m_T > 2 \lambda_t f \sim 2 f$, where $\lambda_t $ is the usual top Yukawa coupling, so that, from eq.~(\ref{delta m_H,little}), 
\beq
\delta m_{H,top}^2(LH) \gsim  1.2 f^2.
 \label{ m_H,little}
\eeq
Now the minimal value that $f$ can attain without undoing the EWPT is very much model dependent \cite{limit-on-littleHiggs, Perelstein:2003wd}. However, taking $f > 1$ TeV as a  minimum limit  in suitably designed models gives already quite a big contribution to the Higgs mass, that has to be cancelled away. I would  care less about this problem if the model could be tested at the LHC even for $f$ significantly greater than 1 TeV, but this may not be the case \cite{Burdman:2002ns, Perelstein:2003wd}. And I do not see any strong reason for insisting on $f \sim 1$ TeV, given that quite a significant cancellation needs anyhow to be swallowed.

Needless to say, in spite of this warning, I believe that possible signals of the "little Higgs" should be carefully looked for.

\section{Relating $G_F$ to  an extra dimension at the weak scale}

\subsection{Motivations and the basic setup}
\label{motivations}

To contemplate the possibility that there exist one or more extra space dimensions may  undoubtedly give a new twist to several of the basic problems in fundamental physics. 
As remarkable examples it suffices to mention the possible connection between the weakness of gravity and a large extra dimension \cite{Arkani-Hamed:1998rs} or the interpretation of the "big" hierarchy, the one between $M_W$ and $M_{Pl}$, as a gravitational blue-shift by a warp factor with a non trivial dependence on a 5th coordinate \cite{Randall:1999ee}. Here I am specifically concerned with the EWSB problem and with the proposition that the Fermi scale be related to the inverse radius of a compactified extra dimension \cite{Antoniadis:1998sd, DPQ, Barbieri:2000vh}. Even this qualification, however, leaves open several possibilities whose description goes beyond the scope of these lectures. In the following I concentrate on a proposal that aims at pushing as far as possible the calculability of the Higgs potential. As it should be clear by now, I look for an optimal solution of the "little hierarchy" problem. To the least, I expect that this will maximize the chances of having this solution tested at the LHC.

The aspect of extra-dimensional theories that interests me most and could have phenomenological applications is the possibility, intrinsically extra-dimensional, of breaking symmetries by boundary conditions on the various fields at the borders of the extra dimension(s). It is in fact even possible to break the same electroweak symmetry in such a way \cite{Csaki:2003dt}, but it is doubtful whether one would improve over 4-dimensional (4D) theories \cite{Barbieri:2003pr}. On the contrary, breaking supersymmetry by boundary conditions on an extra dimension \cite{Scherk:1978ta} looks more promising for reasons that will appear shortly.

All the SM fields are meant to depend also on an extra coordinate $y$, $\Psi (x_{\mu}, y)$, ranging from $0$ to $L\equiv \pi R/ 2$, and are extended to incorporate supersymmetry in the full 5D space. The physics will depend upon the conditions given to the various fields at the boundaries of the $y$-segment $(0, \pi R/ 2)$. Take $(even, even)$ boundary conditions for all the SM fields\footnote{The $y$-segment $(0, L)$ is fictitiously extended at $y<0$ and $y>L$ and the equations of motion are meant to be solved in the whole covering space.}, so that each of them has a constant mode in $y$, with vanishing momentum in the fifth direction, which would otherwise appear as a mass term from the 4D point of view. These "zero modes" are massless. On the contrary, give $(even, odd)$, $(odd, even)$ or $(odd, odd)$ boundary conditions to the extra fields implied by Poincare invariance and supersymmetry in 5D, consistently with the parities at the two boundaries. In this way all these extra fields give massive excitations: from the 4D point of view all their modes have masses which, in the simplest case, are multiple integers of $1/R$. 

This simple construction \cite {Barbieri:2000vh} has a remarkable property. While supersymmetry is manifestly broken in a global way, since there is no supersymmetry in the spectrum - which makes it closer to experiment -, supersymmetry is locally unbroken. This will be spelled out more clearly in the next Section, but is already now intuitively clear. At each boundary all fields can be grouped into doublets of 4D N=1 supersymmetry (one boson and one fermion) with the same boundary conditions, even or odd, inside each supermultiplet. This is unavoidable, given the parities at the two boundaries. On the other hand, the N=1 supermultiplets at the two boundaries are different from each others because of our choice of boundary conditions. At every point of the segment $(0, \pi R/ 2)$ there is at least one 4D supersymmetry, but this supersymmetry changes from point to point since it is different at the two boundaries. This is the basis for the finiteness properties of the loop calculations that have been performed and we are about to describe. All the divergent \emph{local} counterterms have to respect this residual symmetry. The change of supersymmetry at the two boundaries - and, as such, of the corresponding supermultiplets - also allows to describe the Yukawa couplings of the up and down quarks to a single Higgs field as in the SM and unlike in the MSSM, which is not a negligible simplification.

\subsection{The $5D \rightarrow 4D$ projection and the residual symmetries}
\label{setup}

To realize the setup outlined above the fields, organized in N=1 supermultiplets in 5D, 
are the gauge fields $(A_{\mu}, \lambda_1, \lambda_2, \phi_{\Sigma})$, the matter fields $(\psi_M,
   \phi_M, \psi_M^c, \phi_M^c)$ and the Higgs fields
   $(\phi_H,\psi_H,\phi_H^c,\psi_H^c)$. The indices of the $SU(3)XSU(2)XU(1)$ gauge group are left understood, as the flavor indices for matter fields. The boundary conditions assigned to each of these fields, $(+ \equiv even, - \equiv odd)$, as already indicated, are given in Table \ref{tab:kktower}. Only one Higgs supermultiplet (in 5D) appears in Table 
 \ref{tab:kktower}, since only one Higgs scalar will get a vev. The model defined in this way has no gauge anomaly (See App. A). The opportunity to introduce a second Higgs supermultiplet, without a vev, to cancel a Fayet Iliopoulos (FI) term (See App. B) will be discussed later on in Sect. \ref{Higgs-in-det}.

\begin{table}
  \centering
    \begin{tabular}{|c|c|c|c|}
\hline
\hline
      $\psi_M$, $\phi_H$, $A^{\mu}$ &  $\phi_M$, $\psi_H$, $\lambda_1$ &
      $\phi^c_M$, $\psi^c_H$, $\lambda_2$ &  $\psi^c_M$,
      $\phi^c_H$, $\phi_{\Sigma}$ \\
\hline
    $ (+,+) $ & $ (+,-) $ & $ (-,+) $ & $ (-,-) $ \\
\hline
\hline
    \end{tabular}

  \caption{Boundary conditions for
   gauge, matter and Higgs fields at $y=0$ and $y=L$.}
  \label{tab:kktower}
\end{table}

 The supersymmetry transformations for the vector fields, splitted in $A_{\mu}$, $\mu = 1, 2, 3, 4$, and $A_5$, are, as an example, 
 \beq
\delta A_{\mu} = i \xi_i^+ \sigma_{\mu} \lambda_i + h.c., \, \, \, 
\delta A_{5} = \epsilon^{i j} \xi_i \lambda_j+ h.c.,
 \label{ susytrans}
\eeq
where $\xi_i$, ($i, j = 1, 2$) are the two spinorial transformation parameters associated with the 2 supersymmetries in 4D implied by the N=1 supersymmetry in 5D. It is manifest from eqs. (\ref{ susytrans}) that the boundary conditions in Table \ref{tab:kktower} are, in general, not compatible with the supersymmetry transformations, since fields with different boundary conditions transform into each other under (\ref{ susytrans}). Hence we cannot expect that supersymmetry remains unbroken. Nevertheless the consistency between boundary conditions and supersymmetry transformations is kept if the $y$-dependent $\xi_i$ are also restricted to satisfy appropriate boundary conditions, i.e. $(+, -)$ and $(-, +)$ respectively for $\xi_1$ and $\xi_2$.  This defines the residual supersymmetry that remains intact after the $5D \rightarrow 4D$ projection, as alluded to in the previous Section. The same restricted supersymmetry transformations determine the appropriate supermultiplets at the two boundaries. For instance, the two N=1 supermultiplets that contain the Higgs field $\phi_H$ are $\widehat{H} = ( \phi_H, \psi_H)$ at $y=0$ and $\widehat{H_c'} = ( \phi_H^+, \psi_H^c)$ at $y=L$. Accordingly, the total 5D Lagrangian is
\begin{equation}
{\cal L} = {\cal L}_5 + \delta \left(y \right) {\cal L}_4 + \delta
\left( y-\frac{\pi R}{2}\right) {\cal L}_4',
\label{eq:lagrangian}
\end{equation}
where ${\cal L}_5$, the so called "bulk" Lagrangian, respects a full N=1 supersymmetry in 5D, whereas ${\cal L}_4$ and ${\cal L}_4'$ are 4D Lagrangians invariant under the (different) relevant supersymmetries. Among other things, ${\cal L}_4$ and ${\cal L}_4'$ will contain, the first, the Yukawa coupling of the up quarks to $\widehat{H}$ and, the second, the Yukawa couplings of the down quarks and of the leptons to $\widehat{H_c'}$. These couplings could not be placed anywhere else without spoiling the residual supersymmetry.

\subsection{Spectrum of the Kaluza Klein modes}
\label{KKspectrum}

\begin{figure}
\centerline{\epsfxsize=.5\textwidth \epsfbox{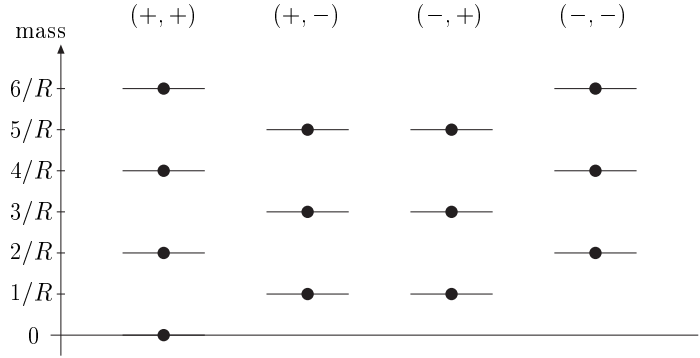}}
\caption{Tree-level KK mass spectrum of a multiplet (vector,
matter or Higgs) with the  boundary conditions as indicated.}
\label{fig:spectrum}
\end{figure}

The solution of the free equations of motion, with the assigned boundary conditions, gives the spectrum of the various modes. This spectrum is particularly simple when one ignores possible kinetic terms localized at the boundaries or gauge invariant "bulk" masses for the matter (or Higgs) supermultiplets. Corresponding to the wave functions 
\begin{equation}
 (+, +) \, : \,  \cos \frac{2 n}{R} y ; \, \, \, 
 (+, -) \, : \,  \cos \frac{2 n + 1}{R} y ; \, \, \,
 (-, +) \, : \, \sin \frac{2 n + 1}{R} y ; \, \, \,
 (-, -) \, : \,  \sin \frac{2 n}{R} y,
\label{wavefunc}
\end{equation}
the spectrum is shown in Fig. \ref{fig:spectrum}. 

Of special interest for the following is the deformation of the spectrum of matter supermultiplets, and of the corresponding wave functions, in presence of a bulk mass term, consistent both with gauge invariance and with the residual supersymmetry. This mass term, in general, can have a $y$-dependence and, for consistency with the parities at the two boundaries, must be $(odd, odd)$. Considering a constant mass term $M$ inside the segment $(0, L)$, all the KK modes become heavier, with a mass growing like $M$ for $M R>> 1$, except for the fermionic zero mode, which remains massless, and one of the two lightest states in the towers of scalars, a sfermion, whose mass decreases from $1/R$ and tends asymptotically to zero as $M R$ increases \cite{Barbieri:2002uk}. The wave functions are modified accordingly. For the massless fermion, taking, e.g., $M > 0$, the corresponding wave function is
 \begin{equation}
 \xi_0(y) = [ \frac{2}{M} (1 - \exp{(-\pi M R)})]^{-\frac{1}{2}} \exp{(-M y)} ,
\label{wavefunc0}
\end{equation}
giving rise to a partial localization towards one of the boundaries ($y = 0$ for $M > 0$), as it happens similarly for the lightest scalar. The physical interpretation is the following: when M becomes larger than $1/R$, supersymmetry is progressively recovered in the spectrum, with a light quasi-supersymmetric multiplet at one boundary and a tower of massive supermultiplets in the bulk. From a theoretical point of view, the bulk mass $M$ is a new parameter, but with a special status: it does not undergo any renormalization.

\subsection{A first attempt: the Constrained Standard Model}
\label{CSM}

\begin{figure}
\centerline{\epsfxsize=.5\textwidth \epsfbox{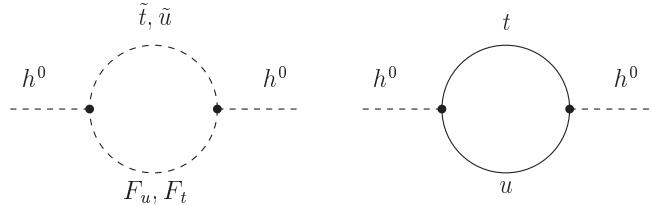}}
\caption{One-loop irreducible diagrams contributing to the mass squared of the
Higgs boson.} 
\label{fig:1loop-diagrams}
\end{figure}

What happens of the EWSB with the simplest setup described in Sect. \ref{setup} ignoring, for the time being, bulk masses and boundary kinetic terms? Needless to say, we are mostly interested in the loop corrections to the Higgs mass from the towers of top-stop states. There are several ways of doing this calculation, either involving a sum over the towers of intermediate KK states \cite{Barbieri:2000vh}  or by working with propagators in mixed $(p_{\mu}, y)$ space for the different components of the superfields \cite{Smith}. With a notation appropriate to this second case, the relevant diagrams are shown in Fig. \ref{fig:1loop-diagrams}, where $h^0(x)$ is the $y$-independent component of the Higgs field and $F_u, F_t$ are the auxiliary components of the $\widehat{u}$ and $\widehat{t}$ supermultiplets, which we have not eliminated. 

The result can best be seen by means of the function  $\Delta m_{H, top}^2(E) $ introduced in Sect. \ref{hierarchy}. It is \cite {Barbieri:2000vh, Smith} 

\begin{eqnarray}
\Delta m_{H, top}^2(E) &=&  \frac{3 \sqrt{2}}{4 \, R^2} G_F m_t^2
\, x^4 \left[ \coth^2 \left(\frac{\pi x}{2}\right) - \tanh^2
\left(\frac{\pi x}{2}\right) \right] 
\label{eq:corr-m2-result}
\end{eqnarray}
where  $x = E R$.
For $E << 1/R$, $ \Delta m_{H, top}^2(E)$ goes to the SM expression, as it should, whereas in the opposite limit, above the compactification scale, $1/R$, $ \Delta m_{H, top}^2(E)$ gets exponentially dumped to
\begin{eqnarray}
\Delta m_{H, top}^2(E) \sim  6 \sqrt{2} G_F m_t^2 E^4 R^2 \exp{(- \pi E R)}.
\label{eq:corr-m2-approx}
\end{eqnarray}
In this way a finite $\delta m_{H,top}^2$ is obtained 
 \beq
\delta m_{H,top}^2(CSM) =\frac{63 \zeta(3)}{\sqrt{2} \pi^4} \frac{G_F m_t^2}{R^2} = \frac{0.19}{R^2},
 \label{ m_H,CSM}
\eeq
where $\zeta(3)=1.20$. I find this an interesting step forward relative to eq. (\ref{delta m_H,top}) or even to eq. (\ref{delta m_H,SS}), which justifies a further exploration of the general idea.

It is possible to go over a complete calculation of the Higgs potential in this simplest case where the top has a flat wave function in $y$. This requires a straightforward extension of the previous calculation to the entire Higgs potential \cite{Barbieri:2000vh}. Such a calculation is relevant both because of the corrections to the quartic Higgs coupling and because of the higher order terms $(h^0)^4 (h^0 R)^n$ \footnote{ With a single Higgs supermultiplet, a quadratically divergent tadpole term in the auxiliary field of the hypercharge vector supermultiplet, a FI term (See App. B) appears \cite{Ghilencea:2001bw}. However, for a sensible cut-off, as determined below in Sect. \ref{where-is-the-cut}, this term is of little numerical significance to the Higgs potential \cite{Barbieri:2001cz}. Alternatively (see Sect. \ref{Higgs-in-det}) a second Higgs supermultiplet can be introduced which cancels the divergent FI term and is prevented from getting a vev by a bulk mass, small relative to $1/R$.}. This corresponds to the minimal implementation of the idea, which we have called Constrained SM \cite{Barbieri:2000vh}, since the relevant part of the full Higgs potential involves a single parameter, $R$, instead of two, as in the SM. Consequently, given the physical value of the Fermi scale, the Higgs mass is determined to be about 130 GeV and the inverse radius is in the $400$ GeV range. Note in particular that, although eq. (\ref{ m_H,CSM}) gives the dominant term in the curvature of the Higgs potential at zero field, the relation between the Higgs mass and $1/R$ cannot be simply read from eq. (\ref{ m_H,CSM}). The curvature of the potential has the standard relation with the Higgs mass only when the powers of $h^0$ higher than four in the potential are unimportant, which is not the case here.

\subsection{Localizing the top: the physical picture}
\label{top-local}

The properties of the top wave function in the fifth dimension are crucial to the numerical determination of the EWSB parameters \cite{Barbieri:2002uk, Marti:2002ar}. In the CSM the top wave function is $y$-independent. As discussed below in Sect. \ref{constraints}, consistency with the EWPT may require the top to be partly localized in $y$, which introduces an extra parameter. The Higgs mass becomes then a function of $1/R$ which is, in turn, not determined \cite{Barbieri:2002sw, Barbieri:2003uk}. 

As already mentioned, in the CSM the one loop top correction to the Higgs mass squared dominates over the gauge corrections, which contribute as \cite{Antoniadis:1998sd}
 \beq
\delta m_{H,ew}^2(CSM) = - \frac{7 \zeta(3) (3 g^2 + g'^2)}{8 \pi^4 R^2} = - 1.5 
\frac{10^{-2}}{R^2},
 \label{ m_H,ew}
\eeq
i.e. one order of magnitude less than eq. (\ref{ m_H,CSM}) and with opposite sign. Note that the relative size between the top and the electroweak contributions is approximately maintained as in the  SM  with a universal cut-off, eq. (\ref{delta m_H^2}).

Compared with this situation, what one should expect with a localized top, through a bulk mass term $M$, is intuitively clear. The progressive localization of the top at one boundary, as $M$ increases, makes it feel less and less the global breaking of supersymmetry through appropriate conditions at the two boundaries. Consequently the top contribution to the curvature of the potential, which vanishes in the supersymmetric limit, decreases its numerical significance. Hence a cancellation takes place between the top and the electroweak contributions, which becomes more and more significant as $M$ increases. In turn, for the physical value of the Fermi scale, this results in an increase of $1/R$ from the relatively low value of the CSM. 

To clarify even further the physical picture, it is useful to see what happens in the extreme case of exact localization of the top at the $y = 0$ boundary, by which I mean that the left handed doublet, $\widehat{Q}$,  and the right handed singlet, $\widehat{u}$,  that contain the top quark, only appear as 4D N=1 supermultiplets in ${\cal L}_4$, eq. (\ref{eq:lagrangian}), with no KK towers. In this case, no supersymmetry breaking enters the diagrams of Fig. \ref{fig:1loop-diagrams}, which do not contribute, therefore, to the curvature of the potential. To see a supersymmetry breaking effect one needs the exchange of a vector or a Higgs supermultiplet, through any of their components, which only occurs at two loops. An explicit two loop calculation gives in fact the following contribution to the Higgs mass, through the curvature of the potential, (LT for "localized top") \cite{Barbieri:2003uk}
\beq
\delta m_{H, top}^2(LT) =  \frac{3 G_F m_t^2}{\sqrt{2} \pi^2} (m_Q^2 \log \frac{c}{R m_Q} + m_U^2\log \frac{c}{R m_U} ),
 \label{delta m_H,loc top}
\eeq
where 
\beq
  m_U^2  =  \frac{7 \zeta(3)}{24 \pi} \frac{8 \alpha_s +6
    \alpha_t}{L^2},  \, \, \, \, \, 
m^2_Q  =  \frac{7 \zeta(3)}{24 \pi} \frac{8 \alpha_s +3
  \alpha_t}{L^2} 
\label{eq:mstops}
\eeq
are the stop squared masses induced at one loop \cite{DPQ}, $\alpha_S = g_S^2/ 4 \pi$, $\alpha_t = \lambda_t^2/ 4 \pi$ and $c = 1.24$. Note again the finite result, as expected. Note also the connection between eq. (\ref{eq:mstops}), and the MSSM result, eq. (\ref{delta m_H,SS}). Numerically it is 
\beq
\delta m_{H, top}^2(LT) =  1.4 \frac{10^{-2}}{R^2}, 
\label{delta m_H,loc top}
\eeq
quite close to $\delta m_{H,ew}^2(CSM) $, eq. (\ref{ m_H,ew}), up to the opposite sign. The one loop electroweak contribution goes through unchanged from the CSM to the top localized case. The near cancellation between eq. (\ref{ m_H,ew}) and eq. (\ref{delta m_H,loc top}) could be at the origin of a significant increase in $1/R$ from the CSM value of about $400$ GeV without fine tunings in the Higgs potential. In turn, as we shall discuss, this may help addressing the "little hierarchy" problem.

\subsection{The Higgs potential in detail}
\label{Higgs-in-det}

To complete the analysis of EWSB, we need the Higgs potential for an arbitrary top localization, as determined by the continuous mass parameter $M$. There are at least two reasons for doing that, other than the logical one. At least the left handed doublets cannot be exactly localized at $y = 0$ if one wants to use only one Higgs with a nonzero vev to describe both the Yukawa couplings to the up and to the down quarks, necessarily sitting at two different boundaries (See Sect. \ref{setup}). Furthermore, the sum of eq. (\ref{ m_H,ew}) and eq. (\ref{delta m_H,loc top}) gives an overall negative contribution which, literally taken, means no EWSB. 
\begin{figure}
  \centering \includegraphics[width=11cm]{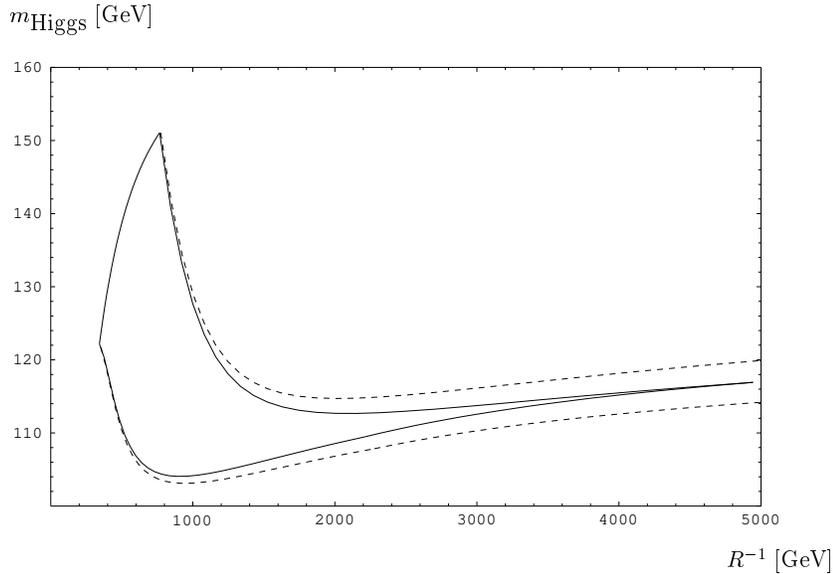}
   \caption{Range of the Higgs mass for a partially localized top as function of $1/R$. The full line is for $m_t^{pole}=174.3\, GeV$. The dotted line is for $m_t^{pole}=174.3\, \pm 5.1 GeV$}
   \label{fig:casomh}
\end{figure}

The Higgs potential with a partial top localization has been studied in Refs. \cite{Barbieri:2002sw, Barbieri:2003uk}. There we consider the $\widehat{Q}$ and $\widehat{u}$ supermultiplets in 5D with an equal bulk mass $M$. We also take two Higgs supermultiplets in 5D with opposite boundary conditions and an equal bulk mass $M_H$, so as to cancel any FI term. A mass term $M_H$ is necessary to stabilize the second Higgs scalar when $M R$ becomes small. For $M R =0$ this is therefore a variant with respect to the CSM, although still with a non localized top. On the contrary, for a sizable $MR$, $M_H$ is progressively negligible.

As already mentioned, the introduction of the parameter $M$ and the near cancellation noticed in the previous Section between the top and the electroweak contributions to the slope of the potential for $M R >> 1$ allow $1/R$ to vary as a free parameter in a wide range, from a few hundreds of GeV for low $M R$ up to several TeV for $M R > 1$, without any significant fine tuning. In Fig \ref{fig:casomh} I show the range of the physical Higgs mass for $1/R$ below 5 TeV and $M_H$ limited by a maximum cancellation in the slope of the potential at about $10 \%$ level. The raise of $m_H$ at low values of $1/R$ is an effect of the KK modes of the top-stop towers. In the upper range of  values for $1/R$, $M_H$ is irrelevant. Values of $1/R$ above 5 TeV would imply a seemingly accidental cancellation between the top and the electroweak contributions by more than a factor of 10. Note that, for $1/R$ above about 1.5 TeV, $m_H$ must be always close to the experimental lower limit of 115 GeV. It must also be said that the calculation of the Higgs potential in Ref. \cite{Barbieri:2003uk}, upon which  Fig. \ref{fig:casomh} is based, is accurate for low $1/R$, below about 1 TeV, where the one loop top contribution dominates, and for high $1/R$, above $3 \div 4$ TeV, where a quasi-localized approximation for the top allows an analytic two loop calculation of the potential. In the transition region a better calculation might be necessary.

\subsection{ Constraints from Electroweak Precision Tests}
\label{constraints}

\subsubsection{ Dominant effects}
\label{dominant-effects}

In general one expects effects on the EWPT from three different sources:
\begin{itemize}
\item[-]
Tree level effects from exchanges of heavy KK modes;
\item[-]
Tree level effects from supersymmetric operators localized on the boundaries;
\item[-]
Calculable  effects from radiatively-generated non-supersymmetric operators.
\end{itemize}

In the following, for concreteness, I do not consider any localization parameter for the first two generations of matter\footnote{When the third generation is localized, this is a source of flavor violation, mostly through a difference between the coupling of the KK gluons to the first two generations and to the third \cite{Delgado:1999sv}. If one assumes mixing angles and phases of the down-quark Yukawa coupling matrix comparable to those of the Cabibbo-Kobayashi-Maskawa matrix, the strongest bound arises from the CP-violating  $\epsilon$-parameter in $K$ physics and is about $1/R > 1.5$ TeV.}. In this case, if one neglects the operators on the boundaries, there is no tree level effect on the EWPT from the exchanges of heavy KK modes, because discrete momentum conservation on the 5th direction forbids any coupling bilinear in the SM particles (the zero modes) to a single KK state \cite{Barbieri:2000vh}. There are, on the contrary, calculable loop effects which are there only to the extent that supersymmetry is broken. A contribution to $\epsilon_3$ would be the best example, as there are several other similar effects on observables not traditionally included among the EWPT, but not less important, like $b \rightarrow s + \gamma$ or the muon $g-2$. Explicit calculations done in the last two cases (but not yet for $\epsilon_3$ \cite{MPinprogr}) show that these effects are below the current sensitivity for $1/R$ above $400\div500$ GeV \cite{Barbieri:2001mr}. Let me therefore concentrate on the effects produced by supersymmetric operators localized on the boundaries.

The coefficients of these operators, all with negative dimension in 
mass, are pure parameters, as would be pure parameters the coefficients of
the higher dimensional operators added to the SM Lagrangian. The most important difference
among the two cases is that the SM Lagrangian is perturbative up to very
high energy whereas the physics described by ${\cal L}_5$ in eq. (\ref{eq:lagrangian}) is not. 
As a consequence, while it is consistent to take  arbitrarily small coefficients for 
the higher dimensional operators in the SM, this is not naturally the case for the theory
under consideration. The operators allowed by the symmetries would at least be 
generated by quantum corrections, with maximal strenght of their coefficients,
at the scale where perturbation theory breaks down. Therefore I consider the 
operators in the sector where perturbation theory breaks down first, at the lowest
energy,  i.e. the operators generated by loops involving the top Yukawa coupling, and I require that their coefficients, as
estimated by means of naive dimensional analysis adapted to 5D,
saturate perturbation theory at the scale $\Lambda$. A theory constructed in this way 
is trustable and predictive, although only at energies sufficiently lower than $\Lambda$.

With this assumption, the operators of interest to the EWPT are (the supersymmetric 
extension of) the Higgs kinetic term and (of) the operator $O_H$ in Table \ref{limits}, both 
localized at $y = 0$,
\beq
\delta {\cal L}_4 = Z_H \vert D_{\mu} H \vert ^2 + C_H \vert H^+ D_{\mu} H \vert ^2 + \dots
\label{localized op}
\eeq
where the dots stand for their supersymmetric completion. Note that $H$ in eq. (\ref{localized op}),
although taken at $y = 0$, is a 5D field with a canonically normalized kinetic term in 
${\cal L}_5$. As such, since $H$ has mass dimension $3/2$, $Z_H$ and $C_H$ have 
dimension $-1$ and $-4$ respectively. From naive dimensional analysis, taking into 
account the angular integration factors 
\beq
\int \frac{d^4 p}{(2 \pi)^4} = \frac{1}{16 \pi^2} \int p^3 dp, \,\,\,\, 
\int \frac{d^5 p}{(2 \pi)^5} = \frac{1}{24 \pi^3} \int p^4 dp
\eeq
 in 4D and 5D respectively, one has 
\beq
Z_H \sim \frac{24 \pi^3}{16 \pi^2} \frac{1}{\Lambda}, \,\,\,\,
C_H \sim \frac{(24 \pi^3)^2}{16 \pi^2} \frac{1}{\Lambda^4}.
\eeq

What is the impact of these operators on the EWPT? The answer is immediate for
$O_H$, being
\beq
H = \frac{1}{\sqrt{2 \pi R}} h^0 + \dots,
\eeq
where $h^0$ is the standard Higgs field, canonically normalized in 4D. In this way,
from the 5D coefficient $C_H$, one obtains the 4D coefficient $c_H/ \Lambda^2 =  C_H/(2 \pi R)^2$ in eq. (\ref{gen SM}) and therefore, from eq. (\ref{epsilon_1}) 
\beq
\Delta \epsilon_1 (O_H) \sim \frac{9 \pi^2}{(\Lambda R)^2} \frac{v^2}{\Lambda^2}.
\eeq
The effect of $Z_H$ on the EWPT is more involved since it gives rise both to a mixing 
of the $W$ and the $Z$ with their KK modes and to a nonzero vev of the scalar
partner of the $W$, i.e. an $SU(2)$ triplet, in the 5D vector supermultiplet (generically
denoted by $\phi_{\Sigma}$ in Table \ref{tab:kktower} ). The net overall effect on the EWPT of this term is again only a contribution to the parameter $\epsilon_1$, which can be read  from Ref. \cite{Delgado:1999sv},(but note the different definition of $R$)
\beq
\Delta \epsilon_1 (Z_H) \sim z_H^2 \frac{ \pi^2}{12} m_Z^2 R^2 
\sim \frac{ 3 \pi^2}{64}\frac{m_Z^2}{\Lambda^2},
\eeq
valid for small $z_H = Z_H/(2 \pi R)$.

\subsubsection{Where is the cut-off?}
\label{where-is-the-cut}

To determine the numerical significance of 
these effects on the EWPT, but not only for this reason, one needs to
know where the cut-off is, or rather its connection with the compactification scale
$1/R$. As already said, this is  related to the energy scale at which perturbation theory
breaks down. In the case of a 5D gauge coupling, $g_5$, taking into account  the
angular integration factor as before, perturbation theory is approximately lost at 
$\Lambda_g \sim 24 \pi^3 / g_5^2$, i.e., from the relation between the
4D and the 5D couplings $g_4^2 = g_5^2 / 2 \pi R$, 
\beq
\Lambda_g R \sim \frac{12 \pi^2}{g_4^2}.
\eeq
 Even in the strong gauge sector therefore, $\Lambda_g$ 
 is higher than $1/R$ by about
 two 
 orders of magnitude. When the top quark is non localized, however, perturbation theory
 in its Yukawa coupling at the $y=0$ boundary 
 \beq
\delta {\cal L}_{4, top} = \lambda_{t,5} \, Q u H 
 \eeq
 gets saturated at a lower scale $\Lambda$, since, from naive dimensional analysis,
\beq
  \lambda_{t,5} \sim \frac{(24 \pi^3)^{3/2}}{16 \pi^2} \frac{1}{\Lambda^{3/2}}
 \eeq 
  and $\lambda_t =  \lambda_{t,5}/ (2 \pi R)^{3/2}$, so that
  \beq
  \Lambda R \sim \frac{4.5}{\lambda_t^{2/3}} \sim 4.5.
  \eeq
  With the progressive localization of the top,  $\Lambda R$  increases up to
$ \Lambda  R \sim 12 \pi^2 / \lambda_t^2$ for an exactly localized top. 
  
We can now get back to the estimate of the effects on the EWPT of 
  $\Delta \epsilon_1 (Z_H)$ and $\Delta \epsilon_1 (O_H) $.  We have
  \beq
  \Delta \epsilon_1 (Z_H) \sim 2 \cdot 10^{-4} (\frac{4.5}{\Lambda R})^2 (R \, TeV)^2
\label{Delta epsilon (ZH)}
  \eeq
  and
  \beq
  \Delta \epsilon_1 (O_H) \sim 6 \cdot 10^{-3} (\frac{4.5}{\Lambda R})^4 (R \, TeV)^2.
\label{Delta epsilon (CH)}
   \eeq
   Note incidentally that $z_H = Z_H/(2 \pi R)$ is indeed small, since it is 
   \beq
   z_H \sim 0.15 (\frac{4.5}{\Lambda R}).
   \eeq
   If we now require $\vert \Delta \epsilon_1 \vert < 2 \cdot 10^{-3}$, 
   the estimate in eq. (\ref{Delta epsilon (ZH)})  allows a low $1/R$, whereas eq. (\ref{Delta epsilon (CH)}), taken at face value, 
   wants $1/R$ above 1.5 TeV. Note in eq. (\ref{Delta epsilon (CH)}), however, the high power of $\Lambda R$,
   which makes the estimate highly uncertain. Note also that $1/R$ increases  from the $400¤$
   GeV range of the CSM by a progressive localization of the top, which, as we said,
  rapidly  increases  $\Lambda R$ and therefore suppresses $\Delta \epsilon_1 (O_H)$ . 

From all this I conclude that values of $1/R$ of about 1 TeV are compatible with the EWPT. Lower values, such as those required by the CSM, seem to need, at the present state of knowledge, an adjustment in $\Delta \epsilon_1$ by about one order of magnitude.
  
\subsection{Spectrum and phenomenology}
\label{spectrum-and-phen}

The most characteristic feature of the spectrum of the superparters is the relative heaviness of gauginos and higgsinos, all approximately degenerate at about $1/R$. The Lightest Supersymmetric Particle (LSP) is therefore a sfermion. In the CSM the LSP is a stop. The relevant spectrum of the CSM is shown in Table \ref{tab:modes-masses} with an estimate of the uncertainties. The lightest stop is stable or quasi stable if a relevant $U(1)_R$ symmetry is slightly broken. 

\begin{table}
\begin{center}
\begin{tabular}{|c||c|c|} 

\hline
\hline
$1/R$                        & $360 \pm 70 $   \\
\hline
$h$                                  & $133 \pm 10 $   \\
\hline
$\tilde t_1 , \, \tilde u_1$          & $210 \pm 20 $  \\
\hline
$\chi^{\pm}, \, \chi^0, \,
\tilde g, \, \tilde q, \, \tilde l$ & $360 \pm 70 $   \\
\hline
$\tilde t_2, \, \tilde u_2$           & $540 \pm 30 $   \\
\hline
$A_1,q_1,l_1,h_1$                    & $720 \pm 140 $  \\
\hline
\end{tabular}
\caption{The particle spectrum and $1/R$ in the CSM. All entries are in GeV.} \label{tab:modes-masses}
\end{center}
\end{table}

Particle localization alters the spectrum of sfermions but does not change the fact that one of them is the LSP.  The masses of the sfermions of charge $Q$
and hypercharge $Y$ are given by
\begin{equation}
   \label{eq:spectrum}
   m^2(Q, Y) =m^2_{\textrm{tree}}+m^2_{\textrm{rad}}(Q, Y) +Y m_Z^2-Q m_W^2
\end{equation}
where $m_{\textrm{tree}}$ is the tree level mass, including the Yukawa
contribution, and $m_{\textrm{rad}}$ is the one loop contribution, as in eq.
(\ref{eq:mstops}). Both
$m_{\textrm{tree}}$ and $m_{\textrm{rad}}$ depend upon the corresponding localization
parameter $MR$ \cite{Barbieri:2002uk, Barbieri:2003uk}. For $MR=0$, $m_{\textrm{tree}}=1/R$, up to the Yukawa term which is important for the stops. For
$MR > 1$, $m_{\textrm{rad}}$ dominates and rapidly approaches the
localized limit given in Table \ref{tm-rad}.

\begin{table}
\begin{center}
\begin{tabular}{|c||c|c|c|c|c|} 

\hline
&  $\widetilde{ Q}$ & $\widetilde {u}$  & $\widetilde{ d }$ & $\widetilde {L}$ & $\widetilde{ e}$ \\ 
 \hline
 $R  \, m_{\textrm{rad}}$ & 0.24 & 0.24 & 0.20 & 0.09  & 0.05\\
  \hline
\end{tabular}
\caption{Radiative masses in units of $1/R$ for the different sfermions. $\widetilde{ Q} = 
(\widetilde{ u}_L, \widetilde{ d}_L) $, $\widetilde{ L} = 
(\widetilde{ \nu}_L, \widetilde{ e}_L) $.}
  \label{tm-rad}
\end{center}
\end{table}
  
Who is the LSP depends therefore upon the localization parameters of the different multiplets. Exact cancellation of the FI term (See App. B) is guaranted by taking equal localization parameters for the quarks within one generation, $M_q$, and, independently, equal localization parameters for the leptons, $M_l$. Remember that these masses are unrenormalized parameters, so it make sense to set them equal. Suppose that only the  third generation quarks are partly localized. Then the LSP is a stop for low $1/R$ and a sbottom for higher $1/R$, due to the difference in $m_{\textrm{rad}}$ ($m_{\textrm{rad}}(\widetilde{ d }) <  m_{\textrm{rad}}(\widetilde{ Q }), m_{\textrm{rad}}(\widetilde{ u })$). The LSP may also be a slepton for a sizable $M_l R$. The difference in $m_{\textrm{rad}}$ ($m_{\textrm{rad}}(\widetilde{e }) <  m_{\textrm{rad}}(\widetilde{ L}))$) makes it definitely more likely to be a charged slepton (in spite of the D-term effects in eq. (\ref{eq:spectrum})).

A stable or metastable charged LSP is a striking phenomenological feature. For definiteness, take it to be a stop (but a sbottom would not make any difference in the experimental signal). By picking up a quark, once pair produced in a high energy collision, it would make any of the two super-hadrons $T^{+}=\widetilde{t} \overline{d}$ or
$T^{0}=\widetilde {t} \overline{u}$ (and their charge conjugates
$T^{-},\overline{T}^{0}$), which should be detectable  as stable particles, since their possible decay into one
another is slow enough to let them both cross the detector. $T^{\pm}$
could appear as a stiff\ charge track with little hadron
calorimeter activity, hitting the muon chambers and
distinguishable from a muon via d$E/$d$x$ and time-of-flight. The
neutral states, on the contrary, could be identified as missing
energy since they would traverse the detector with little
interaction  \cite{Chertok:2001ja}.

\section{Summary and conclusions}
\label{conclus}

The mechanism of EWSB is one of the greatest mysteries in particle physics.
Not the only one, but the one with the greatest chances of being clarified by the
experimentation at the LHC. For the time being, the SM provides a successful 
phenomenological description of EWSB
in terms of a Higgs doublet with a mexican hat potential. This is in fact an 
understatement. The experiments of the ninethies have shown that there must be 
some fundamental truth in the Higgs description of EWSB.
A Higgs boson is very likely to exist. It is likely to be a 
weakly interacting, narrow state. It is pretty likely, although
not certain, that its mass 
is in the $100 \div 
 200$ GeV range. The LHC will tell if this is true or not.

As I said in Sect. \ref{overview}, this goal of the LHC, although clearly important,  is not, however,
the very reason for expecting that the LHC will clarify the mechanism and the dynamics of
EWSB. There are two aspects to this statement. First there is the uneasiness with the
fact that the curvature at zero field of the Higgs potential, crucially negative, is just an input
to the theory without any deeper understanding. Discovering 
the Higgs would not shed light on this problem.
Second,  there is the 
extreme ultraviolet sensitivity of the curvature of the potential: in the SM the Higgs
mass depends on the "Higgs mass spectral function" $\Delta m_H^2$,
 defined in Sect. \ref{hierarchy}, which is rapidly growing with energy and contributes to the Higgs 
 mass
mostly where it cannot be trusted. While the first aspect is generic, the second one is the
basis for thinking that the LHC will reveal new phenomena related to EWSB and not 
included in the SM. At the LHC we should see what dumps $\Delta m_H^2$ at energies
well within the range explored by the machine. In this way we might also understand 
what triggers EWSB by giving the Higgs a negative mass squared.

If this view is right and we are not misguided in a way or another, we are faced with a 
problem. Why haven't we seen yet any manifestation of this dumping mechanism 
anywhere in direct or indirect experiments? The straigthest interpretation of the data, mostly
from the EWPT, points to a significant gap between the Higgs mass and the effective
scale that parametrizes a generic deviation from the SM in terms of higher dimensional
operators, taken maximally symmetric. This is the "little hierarchy"problem. More
precisely the tension is between this scale and the energy at which the dumping
mechanism advocated above should become operative. Finding a theory that dissolves this tension
is to solve the "little hierarchy"problem.  As I have tried to make it
 clear, there are assumptions (and "judgements") in the
line of reasoning that leads to the formulation and the very existence of this problem. To
the point that I am sometimes asked why do I care about it at all. 
To me the answer looks obvious. If these assumptions are valid, the expectation that 
the LHC will clarify the mechanism and the dynamics of
EWSB is well justified. Since they are also reasonable, I prefer to take them seriously.

At least in part, this problem is at the basis of the revival of theoretical interest in EWSB.
Without pretending to be exaustive, I have summarized two attempts at addressing 
this problem, the MSSM and the "little Higgs", and I have described more at lenght a 
third one, based on supersymmetry breaking by extra dimensions. Needless to say,
there is a great physical difference between the first and the two other cases. While the
first can be extended up to very high energy, like the GUT or even the Planck scale, this
in not true for the little Higgs or the extradimensional theories. Quite on the contrary, 
the gap between the dumping scale of $\Delta m_H^2$ and the cut-off of little
Higgs and extradimensional models
 is not more than a decade or so. Is this a step backward? 
Conceptually it may look to be the case. Incidentally this motivates to search for possible
"ultraviolet completions" of these models. On the issue, however, I prefer to keep
an open mind and leave it to the LHC experiments to decide what is relevant and what 
is not. From a theoretical point of view, I find that looking for neat solutions of the
"little hierarchy" problem is a well motivated (and difficult) task. I suspect that the search
will continue.

\section*{Acknowledgements}
\label{sec:acknowledge}
I thank Lawrence Hall, Yasunori Nomura, Riccardo Rattazzi, Alessandro Strumia, Roberto Contino, Michele Papucci, Guido Marandella, Giacomo Cacciapaglia, Marco Cirelli, Paolo Creminelli, Claudio Scrucca, Takemichi Okui, Steven Oliver for many useful discussions. I am indebted to Alessandro Strumia for providing me with Fig. \ref{fig:eps1eps3} and Table \ref{limits}.  It is a pleasure to thank the organizers of this School, especially Dimitri Kazakov and Gerard Smadja.
 This work has been
partially supported by MIUR and by the EU under TMR contract
HPRN-CT-2000-00148.

\appendix

\section{Gauge anomalies}

\label{app:anom}

A gauge theory of fermions in 5D is vector-like. Hence no gauge anomaly is expected. To make contact with phenomenology, however, chiral fermionic zero modes are needed, which is obtained by suitable boundary conditions (BC) on the segment associated with the extra dimension\cite{Dixon:jw, Dixon:1986jc}. In turn, this can generate anomalies localized on the boundaries.

In Sect. \ref{setup} I  have considered two types of BC for the fermions in a 5D supermultiplet:
\begin{itemize}
\item[-]
Matter-like BC: $\Psi ( +, +)$ and $\Psi^c ( -, -)$;
\item[-]
Higgsino-like BC: $\Psi ( +, -)$ and $\Psi^c ( -, +)$.
\end{itemize}
By an explicit calculation one shows that matter-like BC, which produce zero modes, give rise, in general, to anomalies at the two boundaries, equal to each other and to the anomalies of the zero mode fermions. A non anomalous set of zero modes, as in the case of the SM fermions, is therefore the condition for the absence of these anomalies \cite{Arkani-Hamed:2001is}.

On the contrary, higgsino-like BC do not produce massless fermions and, as such, do not give rise to any anomaly. Depending on the regularization procedure, apparent anomalies opposite to each other may be generated at the two boundaries of the $y$-segment \cite{Scrucca:2001eb}, which are, however, cancelled by a suitable Chern-Simons counterterm in 5D \cite{Barbieri:2002ic}.

\section{Fayet Iliopoulos terms}

\label{app:FI}

In supersymmetric theories with a $U(1)$ gauge factor, like the hypercharge of the SM, a quadratically divergent mass term for the scalars of the theory may be generated at one loop. In 4D such a FI term is there if and only if the sum of the $U(1)$-charges of the scalars, $Tr Y$, is non zero. Always in 4D a FI term gives rise in the vacuum to  the breaking either of supersymmetry or of  gauge invariance.

Here we can ask what happens for the scalars of a 5D supersymmetric hypermultiplet, $\phi$ and $\phi^c$, by distinguishing again two types of BC:
\begin{itemize}
\item[-]
Matter-like BC: $\phi ( +, -)$ and $\phi^c ( -, +)$;
\item[-]
Higgs-like BC: $\phi ( +, +)$ and $\phi^c ( -, -)$.
\end{itemize}

In the first case, equal FI terms are generated in general at the two boundaries. The condition for their vanishing is equivalent to the condition $Tr Y = 0$, restricted to the fermionic zero modes in the associated supermultiplets. Again, this condition is fulfilled by the fermions of the SM. In the case of Higgs-like BC, the FI terms at the two boundaries are equal in magnitude and opposite in sign \cite{Ghilencea:2001bw, Barbieri:2002ic}. As such, they are of a particular nature, only possible in 5D. With a 4D superfield notation, their contribution to the action may be written as 
\beq
\delta S_{FI} = \xi \int d^4x \int_0^L dy \int d^4 \theta (\delta_5 V - \Phi - \bar{\Phi}),
\label{SFI}
\eeq
where $V$ is the N=1 vector multiplet and $\Phi$ is the chiral multiplet in the same 5D hypermultiplet. Eq. (\ref{SFI}) and the gauge transformations of the various fields make explicit the gauge invariance of the integrand factor $(\delta_5 V - \Phi - \bar{\Phi})$, unlike what happens for a 4D FI term, where the $\theta$-integration is essential to achieve gauge invariance. This is the basis for the property of a FI term like in eq. (\ref{SFI}) of maintaining both gauge invariance and supersymmetry in the vacuum \cite{Barbieri:2002ic}.

A sufficient condition for avoiding a divergent FI term in the case of 5D multiplets with Higgs-like BC is to have $Tr Y = 0$ for the $(+, +)$ components of the 5D supermultiplets. In this case even a finite FI term is absent at all if the 5D multiplets have equal bulk masses.

\end{document}